\documentstyle[11pt,newpasp,psfig,twoside]{article}
\index{instructions}
\index{guidelines}
\index{Globular Clusters}

\def\edcomment#1{\iffalse\marginpar{\raggedright\sl#1\/}\else\relax\fi}
\reversemarginpar

\begin{document}
\title{A 20\,cm Search for Pulsars in Globular Clusters with Arecibo and the GBT}
\author{J.~W.~T. Hessels$^1$, S.~M. Ransom$^1$, I.~H. Stairs$^2$, V.~M. Kaspi$^1$, P.~C.~C. Freire$^3$, D.~C. Backer$^4$, and D.~R. Lorimer$^5$}

\affil{$^1$McGill, 3600 University St., Montreal, QC H3A 2T8, Canada}
\vspace{-2mm}
\affil{$^2$UBC, 6224 Agricultural Road, Vancouver, BC V6T 1Z1, Canada}
\vspace{-2mm}
\affil{$^3$Arecibo Observatory, HC3 Box 53995, Arecibo, PR 00612, USA}
\vspace{-2mm}
\affil{$^4$UC Berkeley, 601 Campbell Hall, Berkeley, CA 94720, USA}
\vspace{-2mm}
\affil{$^5$U. of Manchester, Jodrell Bank Observatory, Cheshire, SK11 9DL, UK}

\begin{abstract}
We are conducting deep searches for radio pulsations at L-band 
($\sim 20$\,cm) towards more than 30 globular clusters (GCs) using the 305\,m 
Arecibo telescope in Puerto Rico and the 100\,m Green Bank Telescope 
in West Virginia.  
 With roughly three quarters of our search data analyzed, we have discovered 12 new millisecond pulsars (MSPs), 11 of which are in binary systems, and at least three of which eclipse.  We have timing solutions for several of these systems.  
\end{abstract}


There are currently $\sim$76 MSPs known in 23 GCs.  For the last two years we have been searching more than 30 GCs for MSPs with the Berkeley Caltech Pulsar Machine (BCPM) at the GBT and the Wideband Arecibo Pulsar Processor (WAPP) at Arecibo.  
The high time and frequency resolution of these data, along with newly 
developed search algorithms (Ransom, Eikenberry, \& Middleditch 2002; Ransom, Cordes, \& Eikenberry 2003), makes us significantly more sensitive than past surveys to 
sub-millisecond pulsations as well as pulsars in ultra-compact binary systems.
So far we have discovered 12 new MSPs in 6 GCs, 3 of which contained no previously known MSPs.


The newly discovered pulsars are listed in Table 1, along with some of their
 properties.  Only one of the pulsars is isolated.  In an earlier survey of 11 GCs at Arecibo (Anderson 1993) 7 of the 11 pulsars found were isolated (although many of these were found in stack searches).  
Timing observations of all our discoveries are currently underway with Arecibo and the GBT.  

In M13, given the maximum possible cluster  acceleration of $6.0 \times 10^{-18}$ s$^{-1}$              (calculated according to Phinney 1993) and accounting for the measured            proper motion of the cluster and the Galactic gravitational potential, timing  measurements of the four known pulsars show them to be old                    ($\tau_{\rm c} > 2 \times 10^9$ yr) with very low surface magnetic fields 
(B $< 7 \times 10^8$ G).  Similar calculations show that the spin-up of M13D constrains the M/L ratio in the core of the        cluster to be $>$ 2 M$_{\odot}$/L$_{\odot}$ (Ransom et al., in prep.). 

The eclipsing MSP M5C has been identified (D. Pooley, private comm.) in a 
recent Chandra ACIS observation as a soft X-ray source similar to those seen 
in 47 Tuc (Grindlay et al. 2002).  It is also possible that M30A has an X-ray
 association (see Ransom et al. 2003, ApJ submitted, astro-ph/0310347).  

\newpage

\centerline{{Table 1: Newly Discovered MSPs}}
\vspace{-3mm}
\label{tab:psrs}
\begin{center}
\begin{tabular}{lccccc}
\hline \hline
          & $P_{psr}$ & DM & $P_{orbit}$ & $a_1 \sin (i)/c$       & Min $M_2^a$  \\
Pulsar    &      (ms) & (pc cm$^{-3}$) & (hr) & (lt-s) & (M$_{\sun}$) \\
\hline \hline
M30A$^c$  & 11.02 & 25.1 & 4.18    & 0.23  & 0.10       \\
M30B      & 13.0  & 25.1 & $\ga$15 & ?  & $\ga$0.2 \\
M13C$^b$  & 3.722 & 30.1 & ...     & ...   & ...        \\
M13D      & 3.118 & 30.6 & 14.2    & 0.92  & 0.18       \\
M5C$^c$   & 2.484 & 29.3 & 2.08    & 0.057 & 0.038      \\
M5D       & 2.988 & 29.3 & ?       & ?     & ?          \\
M71A$^c$  & 4.888 & 117  & 4.24    & 0.078 & 0.032      \\
M3A       & 2.545 & 26.5 & ?       & ?     & ?          \\
M3B       & 2.389 & 26.3 & 34.0    & 1.9   & 0.20       \\
M3C$^d$   & 2.166 & 26.5 & ?       & ?     & ?          \\
M3D       & 5.443 & 26.3 & ?       & ?     & ?          \\
NGC6749A  & 3.193 & 194  & ?       & ?     & ?          \\
\hline \hline
\end{tabular}
\end{center}

\vspace{-3mm}

\centerline{$^a$Assuming a pulsar mass ($M_1$) of 1.4\,M$_{\sun}$.  $^b$Isolated. }

\centerline{$^c$Shows eclipses.  $^d$Not confirmed.}

\vspace{2mm}

At least 3 of the MSPs (M30A, M71A, and M5C) are in eclipsing systems.  M30A and M5C show eclipse delays from an additional dispersive medium presumably created by the companion's wind.

Although M3 contains 4 new MSPs, they have not been consistently detectable because of scintillation.  We are using multiple WAPP backends, which provide up to 400\,MHz of bandwidth, to increase our detection rate.

Although M30B has been detected only once (owing to the effects of scintillation), orbital analysis of the single detection shows convincing evidence that this pulsar is in a highly relativistic and eccentric (e $\ga 0.5$) binary (see Ransom et al. 2003, ApJ submitted, astro-ph/0310347).

The newest discoveries in the list are M5D and NGC6749A.  M5D was found at Arecibo with the 327\,MHz Gregorian receiver.  We cannot yet be certain that NGC6749A is truly associated with NGC6749; we hope to solidify the interpretation of NGC6749A as a true GC pulsar through timing and hopefully the discovery of other MSPs in this direction with the same DM.  Interestingly, NGC6749 has the lowest concentration ($c=log({\rm r}_t / {\rm r}_c$)) and the second lowest central luminosity density of any GC with known pulsars.

We thank the Canadian Foundation for Innovation and NSERC for support.

\end{document}